\documentclass[twoside,twocolumn,aps,prl,reprint,showpacs,superscriptaddress]{revtex4-1}
\usepackage[latin9]{inputenc}
\usepackage{color}
\usepackage{graphicx}

\makeatletter
 
 \@ifundefined{textcolor}{}
 {%
   \definecolor{BLACK}{gray}{0}
   \definecolor{WHITE}{gray}{1}
   \definecolor{RED}{rgb}{1,0,0}
   \definecolor{GREEN}{rgb}{0,1,0}
   \definecolor{BLUE}{rgb}{0,0,1}
   \definecolor{CYAN}{cmyk}{1,0,0,0}
   \definecolor{MAGENTA}{cmyk}{0,1,0,0}
   \definecolor{YELLOW}{cmyk}{0,0,1,0}
 }

\usepackage[english]{babel}

\makeatother

\begin{document}


\title{Temporal summation in a neuromimetic micropillar laser}

\author{F. Selmi}

\affiliation{Laboratoire de Photonique et de Nanostructures, LPN-CNRS UPR20, Route
de Nozay, 91460 Marcoussis, France}

\author{R. Braive}

\affiliation{Laboratoire de Photonique et de Nanostructures, LPN-CNRS UPR20, Route
de Nozay, 91460 Marcoussis, France}

\altaffiliation{Universit\'e Paris-Diderot, 5 Rue Thomas Mann, 75013 Paris, France}

\selectlanguage{english}%

\author{G. Beaudoin}

\affiliation{Laboratoire de Photonique et de Nanostructures, LPN-CNRS UPR20, Route
de Nozay, 91460 Marcoussis, France}

\author{I. Sagnes}

\affiliation{Laboratoire de Photonique et de Nanostructures, LPN-CNRS UPR20, Route
de Nozay, 91460 Marcoussis, France}

\author{R. Kuszelewicz}

\affiliation{Neurophotonics laboratory,Universit\'e Paris Descartes, 12 rue de l'\'Ecole
de M\'edecine, 75270 Paris Cedex 06, France}

\author{S. Barbay}

\email{sylvain.barbay@lpn.cnrs.fr}

\selectlanguage{english}%

\affiliation{Laboratoire de Photonique et de Nanostructures, LPN-CNRS UPR20, Route
de Nozay, 91460 Marcoussis, France}

\begin{abstract}
Neuromimetic systems are systems mimicking the functionalities or
architecture of biological neurons and may present an alternative
path for efficient computing and information processing. We demonstrate
here experimentally temporal summation in a neuromimetic micropillar
laser with integrated saturable absorber. Temporal summation is the
property of neurons to integrate delayed input stimuli and to respond
by an all-or-none kind of response if the inputs arrive in a sufficiently
small time window. Our system alone may act as a fast optical coincidence detector and
 paves the way to fast photonic spike processing networks.
\end{abstract}

\maketitle

Neuromimetic photonic systems are optical systems that mimic the functionalities
or the architecture of biological neurons, and can represent an alternative path
for computing and processing information very efficiently both
in terms of energy, speed, and robustness versus noise \cite{WoodsNatPhys12,TaitBookChapter14}.

From a functional and basic point of view, a biological neuron can
be thought of as a system than can integrate information from various
stimuli, and respond in an all-or-none fashion if the integrated input
stimuli exceed a certain threshold \cite{KochNatNeuro00}. This latter
property is called excitability and has been experimentally demonstrated
already in many nonlinear semiconductor optical systems, like active
semiconductor cavities with feedback \cite{Giudici97,Wunsche01},
with optical injection \cite{BarlandPRE03,YacomottiPRL06,Goulding07,Beri2010,VanVaerenberghOE12}
or with saturable absorber \cite{BarbayOL11,SelmiPRL14}. The former
property is called temporal summation. Temporal summation refers to
the ability of the system to integrate different, potentially delayed,
presynaptic stimuli and to emit a spike if the integration of the
inputs exceeds the excitable threshold. Since in that case the system
acts as an integrator, it has a time constant and the summation only
takes place if the presynaptic stimuli arrive within in a given time
window. While most of the cortical neurons are integrators \cite{IzhikevichNN01},
note that there also exists so called resonator neurons for which
summation rules depend more on the phase of the input stimuli with
respect to their subthreshold oscillation frequency \cite{IzhikevichNN01,AlexanderOE13}.

In optical systems, the response is in the form of an intensity spike
with a characteristic shape and can be well under the nanosecond timescale
\cite{BarbayOL11,GarbinNatCom15}. Following the demonstration of
excitability in a planar semiconductor laser with integrated saturable
absorber \cite{BarbayOL11}, it has been suggested \cite{NahmiasIEEESTQE13}
that this kind of system could act as a leaky integrate-and-fire neuron,
a model of neuron widespread in neuroscience \cite{IzhikevichIEEETNN04} and optically implemented with
 telecom components in \cite{KravtsovOE11}.

The excitable response of micropillar lasers with integrated saturable absorber
 has been investigated in
\cite{SelmiPRL14} demonstrating the absolute and relative refractory
periods, in complete analogy to biological systems.
In this Letter we investigate the response of a micropillar laser with integrated saturable absorber to sub-threshold stimuli and show
that the system can integrate the stimuli and emit an excitable spike
if the stimuli are close enough in time. This demonstrates temporal summation in this system and its ability
to act as a fast optical coincidence detector \cite{KochNatNeuro00,ShastriCLEO14}, a property usable for optical pattern recognition tasks.
This property is also at the base of some neurocomputational models for e.g. sound localization in
 the barn owl auditory system \cite{PenaILAR10}. It is also
of paramount importance to fabricate optical neuromimetic circuits
by e.g. coupling several micropillar units since, together with excitability,
it demonstrates a fast and compact leaky integrate-and-fire optical neuron.

\begin{figure}
\includegraphics[width=0.8\linewidth]{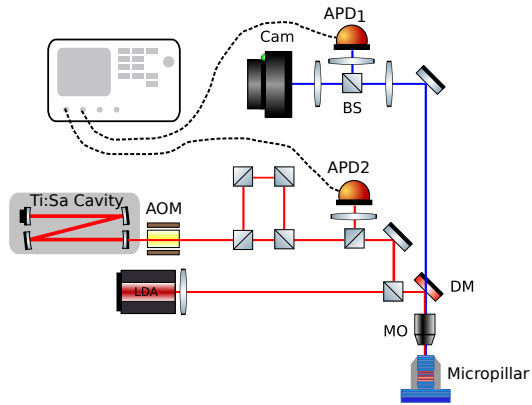} \protect\caption{Experimental setup: LDA, laser diode array; DM, dichroic mirror; BS,
beamsplitter; MO, microscope objective (63$\times$, NA=0,85). Ti:Sa,
mode-locked laser (80ps pulse duration); AOM, pulse picker; APD$_{1,2}$,
avalanche photodiodes; Cam, CMOS camera.}
\label{fig:setup} 
\end{figure}

The experiment consists of a micropillar laser with integrated saturable
absorber of original design described in \cite{ElsassEPJD10,SelmiPRL14}.
The micropillar is optically pumped by a laser diode array emitting
around 800nm and emits light close to cavity resonance designed at
980nm. The active medium consists of two InGaAs/AlGaAs quantum wells
while the saturable absorber medium has only one quantum well. Optical
perturbations are sent to the system thanks to a Ti:Sa model-locked
laser emitting $\sim80$ps duration pulses with a $80$MHz repetition
rate (see Figure \ref{fig:setup}). The rate can be down-sampled thanks
to a pulse picker. A tunable delay line is inserted in the perturbation
path allowing consecutive perturbations with a delay of several hundreds
of picoseconds to several nanoseconds. The pump and perturbation beams
are coupled into the micropillar thanks to a dichroic mirror and a
microscope objective. Two fast avalanche photodetectors ($\mathrm{APD}_{1}$
and $\mathrm{APD}_{2}$) with 10GHz and 4GHz bandwidth respectively
record the input pulses and the response from the micropillar. The
signals are analyzed with a 6GHz oscilloscope.

\begin{figure}
\includegraphics[width=1\linewidth]{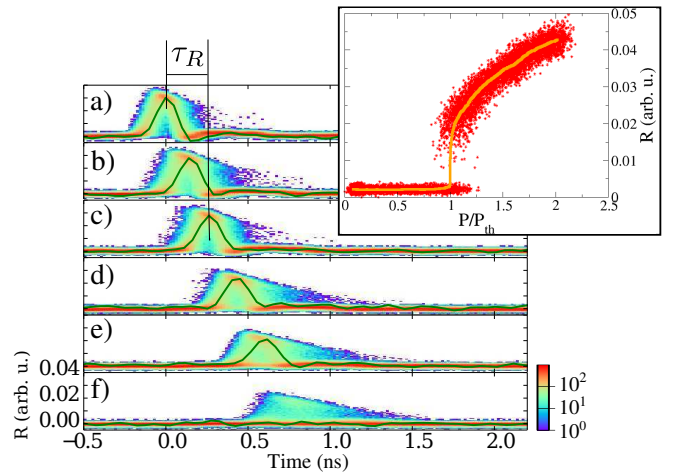} \protect\caption{Experimental traces of the system's response to two incoming, sub-threshold
perturbations for different perturbation delays $\delta$ : a) 210,
b) 320, c) 420, d) 520, e) 610 and f) 700ps. The plots show the statistical
density of points (in log scale) for 10000 different realizations.
On the plots a typical response pulse is shown in green. Inset : excitable
response to a single perturbation. Red stars are the detected response
maxima. Orange is a plot of the median in a sliding window with 500
points.}
\label{fig:traces-sum-resp} 
\end{figure}

\begin{figure}
\includegraphics[width=\linewidth]{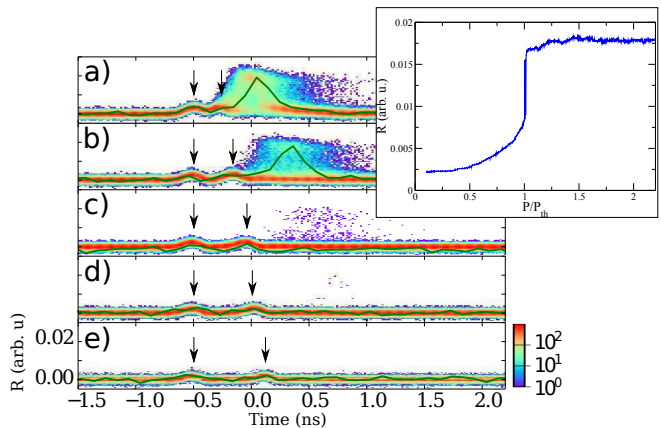}
\caption{Coherent temporal summation with input perturbation wavelength $\lambda=980.471$nm. 
The bias pumping is set to 90\% of the self-pulsing threshold.
The input stimuli 
are set to 44\% and 66\% of the excitability threshold for the direct and delayed pulses respectively.
The plots show the statistical density of points (in log scale) for 10000 different realizations of the input perturbations with delays (a--e) : 220ps, 350ps, 450ps, 540ps and 630ps. On the plots, a typical response pulse is shown in green. The intensity perturbations are indicated by arrows. In inset : excitable response to a single perturbation median-averaged over 500 points.
}
\label{fig:temp-sum-coherent}
\end{figure}

\begin{figure}[t]
\centering{}\includegraphics[width=0.8\linewidth]{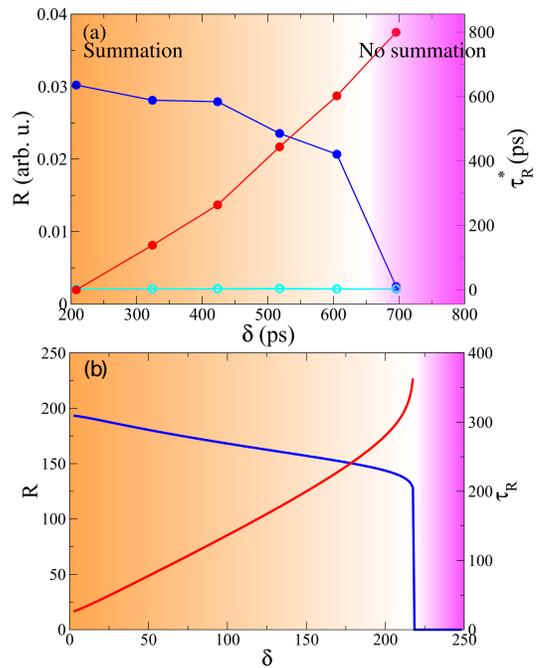} \protect\caption{\label{fig:ampli} (a)Median of the maximum response $R$ (blue) from
experimental data in Fig.\ref{fig:traces-sum-resp} for two consecutive stimuli separated by the delay
$\delta$. The response to each perturbation alone is plotted in light
blue (empty diamond, empty circle respectively). The response delay
(relative to the shortest response delay) $\tau_{R}^{*}$ is shown
in red. (b) Same for model with parameters : $\mu=2.48$, $\mu_{1}=0.43$,
$\mu_{2}=0.43$, $s=10$, $\gamma_{g}=0.005$, $\gamma_{as}=0.01$,
$\gamma=2$, $\eta=1.6$, $\beta=1\times10^{-5}$. The response delay
$\tau_{R}$ is relative to the second perturbation.}
\end{figure}

The micropillar is driven in the excitable regime, 9\% below the self-pulsing
threshold, thus no laser light is emitted. Two consecutive sub-threshold
perturbations are sent onto the micropillar at 797.5nm, close to the
pump wavelength, with a variable delay $\delta$. The first and second
perturbation are set to 74\% and 80\% of the excitable threshold so
that none of each perturbation is able by itself alone to trigger
an excitable response. Since noise is present in the system (either
internal, spontaneous emission noise or external noise sources such
as pump noise or noise in the perturbation amplitude), we record the
response of the system after sending 10000 identical perturbations.
The results are shown on Fig.\ref{fig:traces-sum-resp}. For a large
perturbation delay ($\delta=700\mathrm{ps}$, Fig.\ref{fig:traces-sum-resp}f),
the perturbations rarely adds-up to trigger an excitable response
while for shorter ones, a clear, large amplitude response $R$ is visible.
This means that the system integrates the perturbations which produces
an above-threshold stimulus able to trigger a large, excitable response.
Note that the excitable response, when present, has always the shape
depicted on Fig.\ref{fig:traces-sum-resp}a) when a single event is
plotted (green curve). The average response has a dispersion in time
because of the dynamical delay induced by the noise, which can be
mostly attributed to pump and perturbation noise. In order to quantify
the response we plot in Fig.\ref{fig:ampli}a the median of the response
amplitude $R$ versus delay $\delta$. A clear transition is visible
in the amplitude : for delays below 610ps the perturbations trigger
an excitable response and are thus temporally summed. For larger delays
the summation does not occur anymore. This behavior is in stark contrast
to the case of a gain-switched laser that would respond linearly.
The transition depends on the excitable threshold value and thus on
the bias pump \cite{SelmiPRL14}, and on the amplitude of the incoming
perturbations. 
Summation also occurs for a resonant perturbation at
the cavity resonance close to 980nm, as can be seen on Fig.\ref{fig:temp-sum-coherent}.
In that case, the excitation wavelength is $\lambda=980.471$nm. The bias pump is set to 90\% of 
the self-pulsing threshold in the excitable regime, as can be seen in the inset. 
The two perturbations are set 44\% and 66\% of the excitability threshold for the direct and delayed pulses respectively. They are clearly visible in the recorded traces and are marked by arrows. Temporal summation
occurs for short delays (a), 220ps and to a lesser extent b) 350ps) and is absent for the other delays (c--e) 450, 540 and 630ps). Contrarily to the incoherent excitation case, the excitable curve in inset shows a marked plateau
for stimuli above the excitable threshold. 
This behavior is important in view of cascading several excitable units. 
The fact that temporal summation is possible for incoherent (perturbation on the gain carrier density)
and "coherent" (perturbation on intensity at cavity resonance) adds some flexibility to the system and has
no counterpart in biological systems.

Note also that there is a delay of nonlinear origin
in the response that depends on the summed perturbations (Fig.\ref{fig:ampli}a).
This mechanism is interesting from a neuromimetic point of view since
it provides a natural time-coding mechanism for the amplitudes : for
a large input the delay is short, while it is long for a small (supra-threshold)
summed input.

\begin{figure}[htbp]
\centering{}\includegraphics[width=1\linewidth]{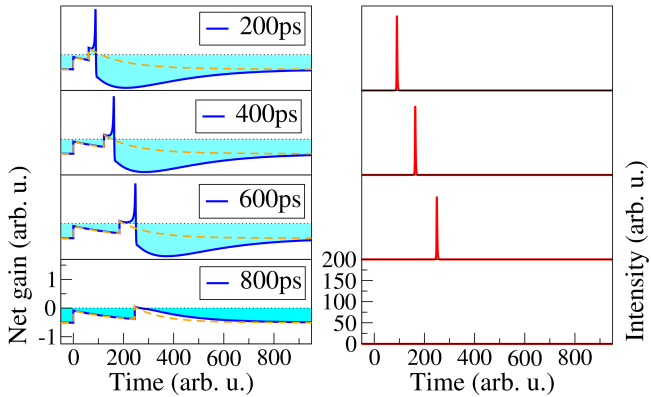} \protect\caption{Numerical simulations of the system Eqs.\ref{eq:yamada} showing the
response to two sub-threshold stimuli on the gain ($\mu_{1,2}$) with
a variable delay $\delta$ : net gain (blue), intensity (red) and
linear solution (Eqs\ref{eq:gain-net}, orange dashed line). Parameters
are the same as in Fig.\ref{fig:ampli}. The excitable threshold is
$\mu_{ex}\simeq0.581$, and $\mu_{1}=\mu_{2}=0.74\mu_{ex}$. The delays
are $\delta=61,\ 122,\ 183\ \mathrm{and}\ 244$ corresponding respectively
to physical delays of 200, 400, 600 and 800ps.}
\label{fig:gainNet} 
\end{figure}

The experimental results are compared to numerical simulations of
the Yamada model with spontaneous emission \cite{DubbeldamPRE99,BarbayOL11}.
This model reads: 
\begin{eqnarray}
\dot{I} & = & (G-Q-1)I+\beta(G+\eta)^{2}\\
\dot{G} & = & \gamma_{g}(\mu-G(1+I))\\
\dot{Q} & = & \gamma_{as}(\gamma-Q(1+sI))\label{eq:yamada}
\end{eqnarray}
. The dynamical variables are the intracavity intensity $I$, the
gain $G$ and the absorption $Q$. Recombination rates of carriers
in the gain and saturable absorber sections are respectively $\gamma_{g}$
and $\gamma_{as}$. Other parameters are $s$ the saturation parameter,
$\gamma$ the linear losses and $\beta$ the spontaneous emission
factor. The term $G+\eta$ is directly proportional to the carrier
density in the gain section. The $\beta$ parameter is small here
and steady state solutions can be expended in power series of $\beta$
\cite{SelmiPRL14}. Let $\{I_{ss},G_{ss},Q_{ss}\}$ be a steady state
solution of Eqs.\ref{eq:yamada}. A simple approach to get an analytic
insight into the dynamics is to consider that the intensity is small
($I_{ss}\propto\beta$) and almost constant as long as a pulse has
not been triggered. Suppose at time $t=0$ a first delta-like sub-threshold
perturbation is sent e.g. on the gain followed by a second perturbation
on at time $t=\delta$ (in units of the photon lifetime in the cavity)
such that $\mu\to\mu+\mu_{1}\delta_{D}(t)+\mu_{2}\delta_{D}(t-\delta)$,
where $\delta_{D}$ is the Dirac delta function. We can then solve
for $G(t)$ such that

\begin{eqnarray}
G(t) & = & G_{ss}+\Pi(t)\mu_{1}\exp\left[\frac{-\gamma_{g}t}{1+I_{ss}}\right]+\nonumber \\
 &  & \Pi(t-\delta)\mu_{2}\exp\left[\frac{-\gamma_{g}\left(t-\delta\right)}{1+I_{ss}}\right]\label{eq:gain-net}
\end{eqnarray}
with $\Pi(t)$ the unit-step function. As noted in \cite{SelmiPRL14},
in a first approximation the net gain $R(t)=G(t)-Q(t)-1$ governs
the triggering of an excitable pulse : a pulse can only be excited
if $R(t)$ exceeds zero for a sufficiently large amount of time. Hence
the final state of the system is controlled by the net gain immediately
after the second perturbation at $t=\delta^{+}$ (always considering
a first, sub-threshold stimulus). Numerical simulations are shown
on Fig.\ref{fig:gainNet}. The net gain is plotted together with the
response to the consecutive stimuli. Simulations of the full system
(Eqs.\ref{eq:yamada}) and of the approximate solution (Eqs.\ref{eq:gain-net})
for the net-gain are shown. The parameters are similar to the one
used in \cite{SelmiPRL14} except the recombinations rates that have
been tuned to match better the experimental results. With the estimated
photon lifetime of the empty cavity being 3.25ps, a reasonable qualitative
agreement is met between the model and the experiment. The critical
delay is found to be $\delta_{c}\simeq218$, that is to say 708ps
in physical units with the estimated photon lifetime. This is in reasonable
agreement with the experimental value in Fig.\ref{fig:ampli}a between
600 and 700ps and for incoherent perturbation conditions similar to those of the experiment:
namely, consecutive perturbations of amplitude 74\% of the single
perturbation excitable threshold $\mu_{ex}$ with the pump at 91\%
of the self-pulsing threshold (corresponding here to the numerical
value of 2.88). The sub-threshold dynamics of the net gain is very
well reproduced by the linear approximation solution until it completely
fails when a pulse is triggered, as expected. For the longest delay
($\delta=244$), when the summed perturbations are sub-threshold,
the approximate net gain deviates from the full solution after the
second perturbation because it passes close to 0 and hence the intensity
dynamics, even failing to trigger an excitable response, has a non-negligible
influence on it. The response and the delay in the response versus
delay between the sub-threshold stimuli are shown in Fig.\ref{fig:ampli}b.
A good agreement is found with the experimental results. The response
abruptly switches to zero when the delay $\delta$ is larger than
a critical delay $\delta_{c}\simeq218$. At this point, the response
delay $\tau_{R}$ diverges as one expects. Note that there is a point
for the response delay $\tau_{R}$ in the experimental curve (Fig.\ref{fig:ampli})
even after the critical delay $\delta_{c}$ because of noise that
is able to trigger a few excitable pulses even for stimuli on average
below the excitable threshold. These events are taken into account
for the determination of $\tau_{R}$. The divergence of the delay
for perturbations close to the excitable threshold is also responsible
for the large dispersion in response times clearly visible in Fig.\ref{fig:traces-sum-resp}d-f)
because a larger number of stimuli will be brought just above the
excitable threshold in these cases. It is interesting to notice that
the final state of the system is different for asymmetric perturbations:
for a balanced perturbation such that $\mu_{1}+\mu_{2}$ is constant,
the highest net gain at time $\delta$ is obtained for asymmetric
perturbations with $\mu_{1}<\mu_{2}$. This slight effect means that
the temporal summation considered here is a non-commutative operation.

\begin{figure}
\includegraphics[width=0.8\columnwidth]{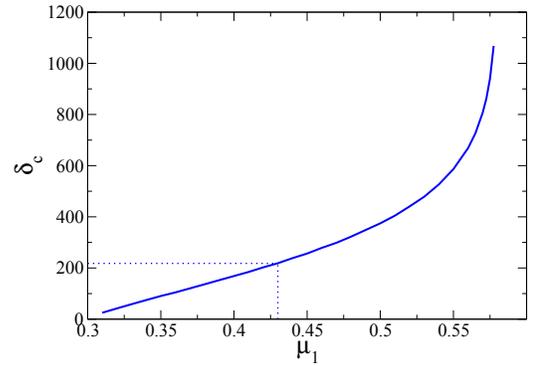}\protect\caption{\label{fig:Critical-delay}Critical delay $\delta_{c}$ versus amplitude
of the two stimuli for $\mu_{1}=\mu_{2}$. Dashed lines correspond
to the case of Fig.\ref{fig:ampli}(b) and $\mu_{1}=\mu_{2}=0.43$.}
\end{figure}

The critical delay is computed for two input stimuli of equal amplitudes
$\mu_{0}$ in Fig.\ref{fig:Critical-delay}. For an amplitude $\mu_{0}<0.37$,
temporal summation never occurs since the summed stimuli are not large
enough to cross the excitable threshold even for zero delay. On the
contrary, for larger stimuli the maximum possible delay giving rise
to an excitable response increases first almost linearly with the
stimulus strength. When a single stimulus is almost able to trigger
an excitable response obviously the maximum delay increases and diverges. 

In conclusion we have studied the effect of consecutive sub-threshold
stimuli on the response of a semiconductor micropillar neuromimetic
system. We have shown that the system can integrate stimuli with a
time constant of the order of several hundreds of picoseconds which
depends on the recombination times of carriers in the active medium
and respond with a macroscopic, excitable pulse if the time delay
between the stimuli is shorter than a critical delay. The critical
delay depends on the excitable threshold which is controllable by 
the amount of pumping of the system. This system thus
simulates the behavior of a leaky integrate-and-fire neuron and is
 particularity suited for building more advanced processing functionalities.
Incoherent and coherent perturbation schemes have been demonstrated, the latter being important
to demonstrate cascadability of such excitable systems.
As was demonstrated in \cite{McCullochBMP43}, a sufficient number
of excitable processing units coupled appropriately is capable of
universal computation. It is however important to keep in mind that
even a single unit has also processing capabilities \cite{KochNatNeuro00}.
In the case of two stimuli, this system behaves as a fast optical
coincidence detector, a functionality that may be of importance in
many applications like pattern recognition in optical signals, in analogy
to biological systems using these property for sound localization
\cite{PenaILAR10}.

\section*{Funding Information}
This work was partially supported by the French Renatech network for
the nanofabrication of the samples.


\begin{thebibliography}{0}%
\makeatletter
\providecommand \@ifxundefined [1]{%
 \@ifx{#1\undefined}
}%
\providecommand \@ifnum [1]{%
 \ifnum #1\expandafter \@firstoftwo
 \else \expandafter \@secondoftwo
 \fi
}%
\providecommand \@ifx [1]{%
 \ifx #1\expandafter \@firstoftwo
 \else \expandafter \@secondoftwo
 \fi
}%
\providecommand \natexlab [1]{#1}%
\providecommand \enquote  [1]{``#1''}%
\providecommand \bibnamefont  [1]{#1}%
\providecommand \bibfnamefont [1]{#1}%
\providecommand \citenamefont [1]{#1}%
\providecommand \href@noop [0]{\@secondoftwo}%
\providecommand \href [0]{\begingroup \@sanitize@url \@href}%
\providecommand \@href[1]{\@@startlink{#1}\@@href}%
\providecommand \@@href[1]{\endgroup#1\@@endlink}%
\providecommand \@sanitize@url [0]{\catcode `\\12\catcode `\$12\catcode
  `\&12\catcode `\#12\catcode `\^12\catcode `\_12\catcode `\%12\relax}%
\providecommand \@@startlink[1]{}%
\providecommand \@@endlink[0]{}%
\providecommand \url  [0]{\begingroup\@sanitize@url \@url }%
\providecommand \@url [1]{\endgroup\@href {#1}{\urlprefix }}%
\providecommand \urlprefix  [0]{URL }%
\providecommand \Eprint [0]{\href }%
\providecommand \doibase [0]{http://dx.doi.org/}%
\providecommand \selectlanguage [0]{\@gobble}%
\providecommand \bibinfo  [0]{\@secondoftwo}%
\providecommand \bibfield  [0]{\@secondoftwo}%
\providecommand \translation [1]{[#1]}%
\providecommand \BibitemOpen [0]{}%
\providecommand \bibitemStop [0]{}%
\providecommand \bibitemNoStop [0]{.\EOS\space}%
\providecommand \EOS [0]{\spacefactor3000\relax}%
\providecommand \BibitemShut  [1]{\csname bibitem#1\endcsname}%
\let\auto@bib@innerbib\@empty
\end{thebibliography}%


\begin{thebibliography}{10}
\newcommand{\enquote}[1]{``#1''}

\bibitem{WoodsNatPhys12}
D.~Woods and T.~J. Naughton, Nat Phys \textbf{8}, 257 (2012).

\bibitem{TaitBookChapter14}
A.~N. Tait, M.~A. Nahmias, Y.~Tian, B.~J. Shastri, and P.~R. Prucnal,
  in \enquote{Nanophotonic Information Physics,} , M.~Naruse, ed., pp. 183 (Springer, 2014).

\bibitem{KochNatNeuro00}
C.~Koch and I.~Segev, Nature Neuroscience \textbf{3}, 1171 (2000).

\bibitem{Giudici97}
M.~Giudici, C.~Green, G.~Giacomelli, U.~Nespolo, and J.~R. Tredicce,
 Phys. Rev. E \textbf{55}, 6414 (1997).

\bibitem{Wunsche01}
H.~J. W\"unsche, O.~Brox, M.~Radziunas, and F.~Henneberger,
 Phys. Rev. Lett. \textbf{88}, 023901 (2001).

\bibitem{BarlandPRE03}
S.~Barland, O.~Piro, M.~Giudici, J.~R. Tredicce, and S.~Balle,
 Phys. Rev. E \textbf{68}, 036209 (2003).

\bibitem{YacomottiPRL06}
A.~M. Yacomotti, P.~Monnier, F.~Raineri, B.~B. Bakir, C.~Seassal, R.~Raj, and
  J.~A. Levenson, Phys. Rev. Lett. \textbf{97}, 143904
  (2006).

\bibitem{Goulding07}
D.~Goulding, S.~P. Hegarty, O.~Rasskazov, S.~Melnik, M.~Hartnett, G.~Greene,
  J.~G. McInerney, D.~Rachinskii, and G.~Huyet, Phys. Rev. Lett.
  \textbf{98}, 153903 (2007).

\bibitem{Beri2010}
S.~Beri, L.~Mashall, L.~Gelens, G.~V. der Sande, G.~Mezosi, M.~Sorel,
  J.~Danckaert, and G.~Verschaffelt, Phys. Lett. A \textbf{374}, 739 (2010).

\bibitem{VanVaerenberghOE12}
T.~V. Vaerenbergh, M.~Fiers, P.~Mechet, T.~Spuesens, R.~Kumar, G.~Morthier,
  B.~Schrauwen, J.~Dambre, and P.~Bienstman, Opt. Express \textbf{20}, 20292 (2012).

\bibitem{BarbayOL11}
S.~Barbay, R.~Kuszelewicz, and A.~M. Yacomotti, Opt. Lett. \textbf{36},
  4476 (2011).

\bibitem{SelmiPRL14}
F.~Selmi, R.~Braive, G.~Beaudoin, I.~Sagnes, R.~Kuszelewicz, and S.~Barbay,
  Phys. Rev. Lett. \textbf{112}, 183902 (2014).

\bibitem{IzhikevichNN01}
E.~M. Izhikevich, Neural Networks
  \textbf{14}, 883 (2001).

\bibitem{AlexanderOE13}
K.~Alexander, T.~V. Vaerenbergh, M.~Fiers, P.~Mechet, J.~Dambre, and
  P.~Bienstman, Opt. Express
  \textbf{21}, 26182 (2013).

\bibitem{GarbinNatCom15}
B.~Garbin, J.~Javaloyes, G.~Tissoni, and S.~Barland, Nat Commun \textbf{6},
5915 (2015).

\bibitem{NahmiasIEEESTQE13}
M.~Nahmias, B.~Shastri, A.~Tait, and P.~Prucnal,
IEEE J. Sel. Topics Quantum Electron. \textbf{19}, 1 (2013).

\bibitem{IzhikevichIEEETNN04}
E.~Izhikevich, IEEE
  Trans. Neural Netw. \textbf{15}, 1063--1070 (2004).

\bibitem{KravtsovOE11}
K.~S. Kravtsov, M.~P. Fok, P.~R. Prucnal, and D.~Rosenbluth, Opt.
  Express \textbf{19}, 2133 (2011).

\bibitem{ShastriCLEO14}
B.~J. Shastri, A.~N. Tait, M.~Nahmias, B.~Wu, and P.~Prucnal,
 in
  \enquote{CLEO: 2014,}  (Optical Society of America, 2014), p. STu3I.5.

\bibitem{PenaILAR10}
J.~L. Pe\~na and W.~M. DeBello, ILAR journal \textbf{51}, 338 (2010).

\bibitem{ElsassEPJD10}
T.~Elsass, K.~Gauthron, G.~Beaudoin, I.~Sagnes, R.~Kuszelewicz, and S.~Barbay,
  Eur. Phys. J. D \textbf{59},
  91 (2010).

\bibitem{DubbeldamPRE99}
J.~L.~A. Dubbeldam, B.~Krauskopf, and D.~Lenstra, Phys. Rev. E
  \textbf{60}, 6580 (1999).

\bibitem{McCullochBMP43}
W.~S. McCulloch and W.~Pitts, The bulletin of mathematical biophysics \textbf{5},
  115 (1943).

\end{thebibliography}
\end{document}